\documentclass{elsart}
\usepackage[latin1]{inputenc} 
\usepackage[german,english]{babel} 
\usepackage{longtable} 
\usepackage{rotating} 

\usepackage{graphics}
 \usepackage{graphicx}
 \usepackage{epsfig}

\usepackage{amssymb}

\begin{document}

\begin{frontmatter}



\title{Elliptic solutions to a generalized BBM equation}

 \author[Osnabrueck]{J. Nickel}
  \address[Osnabrueck]{Department of Physics, University of Osnabrück, Barbarastr. 7, D-49069 Osnabrück, Germany}



\begin{abstract}
An approach is proposed to obtain some exact explicit solutions in
terms of the Weierstrass' elliptic function $\wp$ to a generalized
Benjamin-Bona-Mahony (BBM) equation. Conditions for periodic and
solitary wave like solutions can be expressed compactly in terms
of the invariants of $\wp$.
The approach unifies recently established ad-hoc methods to a certain extent.
Evaluation of a balancing principle simplifies the application of this approach.

\end{abstract}

\end{frontmatter}

\section{Introduction}

The Benjamin-Bona-Mahony equation has been investigated as a
regularized version of the Korteweg-de Vries equation for shallow
water waves \cite{Zhang2001}. It incorporates nonlinear dispersive
and dissipative effects \cite{Wan1997,Wan1998}. In certain
theoretical investigations the equation is superior as a model for
long waves, and the word "regularized" refers to the fact that,
from the standpoint of existence and stability, the equation
offers considerable technical advantages over the Korteweg-de
Vries equation. In addition to shallow water waves, the equation
is applicable to the study of drift waves in plasma or the Rossby
waves in rotating fluids. Under certain conditions, it also
provides a model of one-dimensional transmitted waves. These find
applications in semiconductor devices, optical devices,
etc. \cite{Zhang2001}. -
Apart from these applications solutions to the BBM equation are interesting in and of themselves.\\
Recently, Wazwaz \cite{Waz2006a} has introduced five new ansaetze,
 sinh-cosh-ansatz I - III, a sine-cosine- ansatz and a tanh-sech-method
to find exact solutions to a generalized BBM equation

\begin{equation}\label{BBM.f1}
  u_t + u_x + a\: u^n u_x + u_{xxx} = 0, \: n \geq 1,
\end{equation}

with constant parameter $a$. The ansaetze are direct approaches
and "ad hoc" as a method, but yield (according to \cite{Waz2006a})
new results. Therefore, it seems reasonable to introduce and apply
a further direct method to the above equation. As an interesting
"feature" this method has compact conditions for periodic and
solitary solutions. The power of the method is its ease of use and
the existence of the mentioned conditions. Furthermore, it will be
shown that the sinh-cosh-ansatz I-II, the sine-cosine-ansatz, the
tanh-sech-method and the sinh-cosh-ansatz III for $n = 1$ are
subcases of the
method presented here.\\

The main features of this method are outlined in sec.
\ref{Outline} and applied to the BBM equation in brief. A balancing principle is evaluated in subsec. \ref{Balance}.
Periodic
and solitary solutions to the BBM are presented in subsec.
\ref{BBM}. The relationship between this method and the ansaetze
introduced in \cite{Waz2006a} are shown in sec.
\ref{Relation}. A summary is given in sec. \ref{Con}.

\section{Outline of the method \cite{Schue2006,Nic2006b}}\label{Outline}

The starting point is a nonlinear wave and evolution equation (NLWEE)

\begin{equation}\label{NLWEE.f1}
\tau[ u(\textbf{x}), u'(\textbf{x}), u''(\textbf{x}), ...,
u^{(k)}(\textbf{x}), ..., u^{(m)}(\textbf{x})]
 = t(\psi),
\end{equation}

where $\tau$ is a function of $u(\textbf{x})$ and its partial
derivatives, the independent variable $\bf x$ has n components
$x_j$, $u^{(k)}(\bf{x})$ denotes the collection of mixed
derivative terms of order $k$ and $t(u)$ is a trigonometric
function in $\psi$ or $ t(u) = 0$. For
notational simplicity the independent variables $x$, $t$ will be used in the following.\\
Equation (\ref{NLWEE.f1}) describes a certain dynamical system by
means of a (wave) function $u(\bf{x})$. An travelling wave ansatz

\begin{equation}\label{Out.f0}
u(x,t) \to  g(f(z)) = \sum_{i = 0}^M a_i f(z)^i, \: M \in \mathbb{N},
\end{equation}

where $f$ is supposed to obey the nonlinear differential equation

\begin{equation}\label{Out.f1}
\left( \frac{df(z)}{dz}\right)^2 = \alpha f^4 + 4\beta f^3+6\gamma
f^2+4\delta f+\epsilon\equiv R(f),
\end{equation}

(with real  $\alpha$, $\beta$, $\gamma$, $\delta$, $\epsilon$, $z
= \mu(x - c t)$, $f(z)$) transforms the NLWEE into an polynomial equation $P(f) = 0$.
 Vanishing coefficients in the polynomial equation
$P(f) = 0$ lead to equations which partly determine the
coefficients $\alpha$, $\beta$, $\gamma$, $\delta$, $\epsilon$ in
eq. (\ref{Out.f1}). In general, the coefficients depend on the
structure and parameters of the NLWEE and, finally, on the parameters of the transformation
$\psi \rightarrow g$. The parameter $M$ in eq. (\ref{Out.f0}) can
be derived by balancing ( cf. subsec. \ref{Balance}).

Thus, the problem of finding a solution to the NLWEE is reduced to finding an appropriate
transformation that leads to eq. (\ref{Out.f1}), which, in this
sense, is the basis of the following analysis (for reference
purposes it is called the
''basic equation'' of the associated NLWEE).\\

As is well known \cite{10,10a} the solutions $f(z)$ of
(\ref{Out.f1}) can be expressed in terms of Weierstrass' elliptic
function $\wp$. It reads

\begin{equation}\label{Out.f4}
f(z)= f_0+ \frac { R'(f_0) } {4[\wp(z;g_2,g_3)-\frac{1}{24}
R''(f_0)]},
\end{equation}

where the primes denote differentiation with respect to $f$ and
$f_0$ is a simple root of $R(f)$.

 The invariants $g_2,g_3$ of
Weierstrass' elliptic function $\wp(z;g_2,g_3)$ are related to the
coefficients of $R(f)$ by \cite{11}

\begin{equation}\label{Out.f5}
g_2= \alpha\epsilon-4\beta\delta+3\gamma^2 \quad, \quad
\end{equation}

\begin{equation}\label{Out.f6}
g_3= \alpha\gamma\epsilon+2\beta\gamma\delta-\alpha\delta^2
-\gamma^3-\epsilon\beta^2\quad. \quad
\end{equation}

The invariants and the discriminant (of $\wp$ and $R$ \cite{11})

\begin{equation}\label{Out.f8}
\triangle= g_2^3- 27 g_3^2, \quad
\end{equation}

are suitable to classify the behaviour of $f(z)$. The conditions
\cite{8a}

\begin{equation}\label{Out.f16}
\triangle\neq 0\quad {\mathrm {or}}\quad \triangle=0 \quad, \quad
g_2>0,\quad g_3>0.
\end{equation}

lead to periodic solutions, whereas the conditions

\begin{equation}\label{Out.f9}
\triangle=0,\quad g_2\geq 0\quad ,\quad g_3\leq 0
\end{equation}

are associated with solitary solutions.

If $\triangle=0$, the solution (\ref{Out.f4}) can be simplified
because $\wp(z;g_2,g_3)$ degenerates into trigonometric or
hyperbolic functions \cite[pp. 651--652]{9}.

Thus, in this case, periodic solutions according to eq.
(\ref{Out.f4}) are determined by

\begin{equation}\label{NVE.f7b}
 f(z)= f_0+\frac{R'(f_0)} {4 \left[
-\frac{e_1}{2}-\frac{R''(f_0)}{24}+ \frac{3}{2}e_1\csc^2
(\sqrt{\frac{3}{2}e_1}z)\right]},\quad \triangle = 0,\: g_3>0,
\end{equation}

and solitary wave like solutions by

\begin{eqnarray}
f(z)&=& f_0+\frac{R'(f_0)} {4 \left[ e_1-\frac{R''(f_0)}{24}+3e_1
\mathrm{csch}^2(\sqrt{3e_1}z)\right]},\quad
\triangle = 0,\: g_3<0,\label{NVE.f7c}\\
 f(z)&=& f_0+\frac{6R'(f_0)z^2}{24-R''(f_0)z^2},\quad \triangle = 0,\: g_3=0,\quad  R''(f_0) <
 0,\label{NVE.f7d}
\end{eqnarray}

where $e_1= \sqrt[3]{|g_3|}$ in eq. (\ref{NVE.f7b}) and
$e_1=\frac{1}{2}\sqrt[3]{|g_3|}$ in eq. (\ref{NVE.f7c}).

In general, $f(z)$ (according to eq. (\ref{Out.f4})) is neither
real nor bounded. Conditions for real and bounded solutions $f(z)$
can be obtained by considering the phase diagram of $R(f)$
\cite[p. 15]{6a}. They are denoted as "phase diagram conditions"
(PDC) in the following. Examples of a phase diagram analysis is
given in refs. \cite{Nic2006d,8a,6a}.

\subsection{Balancing principle}\label{Balance}

As mentioned above, the exponent $M$ in eq. (\ref{Out.f0}) is determined by use of a balancing principle. In this subsection a balancing principle
for the approach outlined above is evaluated.

As a consequence of the transformation $u \rightarrow g(f(z))$ (cf. eq. (\ref{Out.f0})), the NLWEE  can be rewritten in terms of $g(f)$
and its derivatives. To determine the leading $M$ exponent

\begin{equation}\label{KSE.f8}
\widetilde{g}(f) = a_0 + a_M f^M
\end{equation}

has to be substituted into the NLWEE in question that leads to an
expression of the form

\begin{eqnarray}
\widetilde{P} (f) &=& 0,\label{KSE.f9}\\
\sqrt{\alpha f^4 + 4 \beta f^3 + 6 \gamma f^2 + 4 \delta f + \epsilon}\:\:
 \widetilde{Q}(f) &=& 0,\label{KSE.f9a}\\
 \widetilde{P} (f) + \sqrt{\alpha f^4 + 4 \beta f^3 + 6 \gamma f^2 + 4 \delta f + \epsilon}\:\:
 \widetilde{Q}(f) &=& 0,\label{KSE.f9b}
\end{eqnarray}

where $\widetilde{P}$ and $\widetilde{Q}$ are polynomials in $f$.
To determine $M$ in ansatz (\ref{Out.f0}) -
according to type (\ref{KSE.f9}), (\ref{KSE.f9a}),
(\ref{KSE.f9b}) - the polynomial equations $\widetilde{P}(f) = 0$, $\widetilde{Q}(f) = 0$
or the equation

\begin{equation}\label{Out.f3}
\widetilde{P}(f)^2 - R(f)\: \widetilde{Q}(f)^2 = 0
\end{equation}

is considered and every two possible highest
exponents are equated. To avoid applying this procedure for every NLWEE anew, the highest possible exponents of the
polynomials that occur for special derivatives in the nonlinear
wave and evolution equation in question have been evaluated.
One obtains from equation (\ref{Out.f1})

\begin{eqnarray}
\frac{d}{dz} &\rightarrow& \sqrt{\alpha f^4 + 4 \beta f^3 + 6
\gamma f^2 + 4 \delta f + \epsilon} \frac{d}{df},\label{Out.f1a}\\
\frac{d^2}{dz^2} &\rightarrow& (2 \alpha f^3 + 6 \beta f^2 + 6
\gamma f + 4 \delta) \frac{d}{df},\label{Out.f1b}\\
\frac{d^3}{dz^3} &\rightarrow& (6 \alpha f^2 + 12 \beta f + 6
\gamma) \sqrt{\alpha f^4 + 4 \beta f^3 + 6 \gamma f^2 + 4 \delta f
+ \epsilon} \frac{d}{df}\label{Out.f1c}\\
&+& (2 \alpha f^3 + 6 \beta f^2 + 6 \gamma f + 4 \delta)
\sqrt{\alpha f^4 + 4 \beta f^3 + 6 \gamma f^2 + 4 \delta f +
\epsilon} \frac{d^2}{df^2}.\nonumber
\end{eqnarray}

Consideration of these relations and eq. (\ref{KSE.f8})
leads to the following list of the highest possible  exponents (HPEs) (cf. tab. \ref{Table1}).

\begin{table}[h]
\begin{tabular}{|l||l|l|l|l||l|l|}\hline
Function & $g$ & $g_z$ & $g_{zz}$ & $g_{zzz}$ & $g^n\: g_z$& \ldots\\\hline
HPE of $\widetilde{P}$ & $M$ & $0$ & $M + 2$ & $0$ & $0$& \ldots\\\hline
HPE of $\widetilde{Q}$ & $0$ & $M-1$ & $0$ & $M + 1$ & M (n + 1) - 1& \ldots\\\hline
\end{tabular}
\caption{\label{Table1} Highest
possible exponent (HPE) of function $\widetilde{g}$ and its derivatives
according to (\ref{KSE.f8}) and equations
(\ref{Out.f1a}) - (\ref{Out.f1c}).}
\end{table}

\subsection{Solutions to the BBM equation}\label{BBM}

Using a transformation\footnote{Balancing $g_{zzz}$ and $g^n\: g_z$ yields $M = \frac{2}{n}$ according to tab. \ref{Table1}.
To simplify the comparison with the ansaetze in \cite{Waz2006a} $M = \frac{1}{n}$ is chosen. It is obvious, that solutions
(\ref{BBM.f6}), (\ref{BBM.f7}) and (\ref{BBM.f8}) can easily be rewritten as $\widetilde{h}(z)^{\frac{2}{n}}$ as expected from the balancing result.}

\begin{eqnarray}\label{BBM.f2}
 u(x,t) &=& g(z), \: z = \mu(x - c t),\nonumber\\
 g(z) &=& h(z)^{\frac{1}{n}},\\
 h(z) &=& \sum_{j = 0}^2 a_j f^j,\nonumber
\end{eqnarray}

where $f$ is supposed to obey eq. (\ref{Out.f1}), leads to a
polynomial equation $P(f) = 0$.
Setting the coefficients equal to
zero yields

\begin{eqnarray}\label{BBM.f3}
\alpha &=& -\frac{a a_2 n^2}{4 (n + 4) (n + 1) \mu^2}, \: \beta =
-\frac{a a_1 n^2}{8 (n + 4) (n + 1) \mu^2}, \nonumber\\
\gamma &=& \frac{n^2 (-3 a (a_1^2 + 4 a_0 a_2) + a_2 (c - 1) (n +
4) (n + 1))}{96 a_2
(n + 4) (n + 1)},\nonumber\\
\delta &=& \frac{a_1 n^2 (a (a_1^2 - 12 a_0 a_2) + a_2 (c - 1) (n
+
4) (n + 1))}{64 a_2^2 (n + 1) (n + 4) \mu^2},\\
\epsilon &=& -\frac{n^2 (a (a_1^4 - 12 a_0 a_1^2 a_2 + 48 a_0^2
a_2^2) + a_2 (a_1^2 - 8 a_0 a_2) (c - 1) (n + 4) (n + 1))}{64
a_2^3 (n + 1) (n + 4) \mu^2},\nonumber\\
a_0 &=& \frac{a_1^2}{4 a_2} \vee a_0 = \frac{a a_1^2 + a_2 (c - 1)
(n + 4) (n + 1)}{4 a a_2}.\nonumber
\end{eqnarray}

According to eqs. (\ref{Out.f5}) - (\ref{Out.f8}) the invariants
of the Weierstrass' elliptic function and the discriminant read

\begin{eqnarray}\label{BBM.f4}
g_2 = \frac{(c - 1)^2 n^4}{3072 \mu^4}, g_3 = -\frac{(c - 1)^3
n^6}{884736 \mu^6}, \triangle = 0 \: \textrm{if}\: a_0 =
\frac{a_1^2}{4 a_2};
\end{eqnarray}


\begin{eqnarray}\label{BBM.f5}
g_2 &=& \frac{(c - 1)^2 n^4}{192 \mu^4}, g_3 = -\frac{(c - 1)^3
n^6}{13824 \mu^6}, \triangle = 0\nonumber\\
 \: \textrm{if}\: a_0 &=&
\frac{a a_1^2 + a_2 (c - 1) (n + 4) (n + 1)}{4 a a_2}.
\end{eqnarray}


As can be seen from eqs. (\ref{BBM.f4}) and (\ref{BBM.f5}),
periodic solutions are given if $c < 1$, whereas $c
> 1$ leads to solitary solutions, according to conditions
(\ref{Out.f16}), (\ref{Out.f9}).

If $a_0 = \frac{a_1^2}{4 a_2}$, periodic and solitary solutions
can be evaluated according to eqs. (\ref{BBM.f2}), (\ref{NVE.f7b})
and (\ref{NVE.f7c})\footnote{The case $g_2 = g_3 = 0$ is neglected
here, because according to the PDC there exist no bounded
solutions.}, respectively. The periodic solution reads

\begin{equation}\label{BBM.f6}
  u(x,t) = \left\{\frac{(c - 1) (n + 4) (n + 1)\sec^2[\frac{n}{4} (x - c t) \sqrt{(1 - c)}]}{4
a}\right\}^{\frac{1}{n}}.
\end{equation}

Solitary solutions are given by

\begin{equation}\label{BBM.f7}
  u(x,t) = \left\{\frac{(c - 1) (n + 4) (n + 1)\textrm{sech}^2[\frac{n}{4} (x - c t) \sqrt{(c - 1)}]}{4
a}\right\}^{\frac{1}{n}},
\end{equation}

\begin{equation}\label{BBM.f8}
u(x,t) = \left(\frac{\{a a_1^3 + 2 a_2 \sqrt{2 a a_1^2 (c - 1) (n
+ 2) (n + 1)}\textrm{sech}[\frac{n}{2} (x - c t) \sqrt{(c - 1)
}]\}^2}{16 a^2 a_1^4 a_2}\right)^{\frac{1}{n}}.
\end{equation}

If $a_0 = \frac{a a_1^2 + a_2 (c - 1) (n + 4) (n + 1)}{4 a a_2}$
the basic equation has two double roots. Therefore, the general
solution to eq. (\ref{Out.f1}) \cite{10,10a}

\begin{equation}\label{BBM.f8a}
f(z)=f_0+ \frac {\sqrt{R(f_0)}
   \frac{d\wp(z;g_2,g_3)}{dz}+
   \frac{1}{2} R'(f_0) [\wp(z;g_2,g_3)-
   \frac{1}{24} R''(f_0)]+
   \frac{1}{24}R(f_0) R'''(f_0)}
{2 [\wp(z;g_2,g_3)-\frac{1}{24}R''(f_0)]^2-
    \frac{1}{48}R(f_0) R''''(f_0)  },
\end{equation}

where $f_0$ is any constant, not necessarily a simple root of
$R(f)$, has to be considered. Evaluation yields exactly solutions
(\ref{BBM.f6}) and (\ref{BBM.f7}).\\

Equations (\ref{BBM.f6}), (\ref{BBM.f7}) and (\ref{BBM.f8})
present periodic and solitary solutions to the BBM equation.
Because of the generalized ansatz there are no
restrictions for $a_1$ and $a_2$. This may be advantageous to
adapt the theoretical results to a physical problem that is
modelled by the BBM equation.

\section{Relationship between the approach above to some other ad-hoc methods}\label{Relation}

Recently, Wazwaz has introduced five ansaetze to solve the BBM
equation \cite{Waz2006a}. Comparing these ansaetze with the
approach outlined above, shows, that four of the ansaetze occur as
special cases of the approach here. The $\sinh-\cosh$-ansatz III
can only be classified as a special case  if $n = 1$. It is
remarkable, that this classification as "special cases" does not
depend on the nonlinear equation in question, but is a
characteristic of the approach outlined above. Therefore, features
like the balancing principle, conditions for periodic and solitary
wave solutions and the PDC can also be applied to the ansaetze
given in \cite{Waz2006a}.
This may be of interest because these ansaetze have also been applied succesfully to the Boussinesq equation \cite{Waz2006b}.\\

As mentioned above the exponent $M$ in eq. (\ref{Out.f0}) can
always be obtained by balancing\footnote{In ref. \cite{Waz2006a} the exponent in the $\sinh-\cosh$-ansaetze I-III
is directly given as $\frac{1}{n}$. This result is obtained by balancing here.}. Therefore, the
base of the ansaetze in \cite{Waz2006a} has to be compared with
eq. (\ref{Out.f1}). The comparison is as follows: An ansatz $f(z)$
according to \cite{Waz2006a} with $z = x - c t$ is considered. The
function $f$ is differentiated with respect to $z$ and $\left(
\frac{df(z)}{dz}\right)^2$ is rewritten in terms of $f$. Comparing
this expression with eq. (\ref{Out.f1}) leads to the relationship
between the parameters in the ansatz and the parameters in the
basic equation. The results are presented in tab. \ref{TableA1}.\\

It should be pointed out that the following classification of the
ansaetze as subcases of a generalized ansatz (\ref{Out.f0}),
(\ref{Out.f1}) has several advantages: Firstly, it is shown that
ad-hoc methods can be unified to a certain extent. Secondly,
periodic and solitary solutions can be deduced systematically
according to conditions (\ref{Out.f16}) and (\ref{Out.f9}).
Thirdly, certain procedures, e.~g. balancing, has only to be done
once and not for every ansatz anew. Finally, some of the remarks
concerning solutions evaluated in ref. \cite{Waz2006a} (confered
to as "W") can be approved from another point of view: The remark,
that solutions given by eqs. (W20), (W21) are consistent with
solutions (W18), (W19) can be approved by considering tab.
\ref{TableA1}. For $\lambda = \pm 1$ in the cosh I -ansatz and
$\lambda = \pm i$ in the sinh I-ansatz the parameters of the basic
equation are equal ($\alpha = 0$ in both cases), so that the same
solutions have to be expected.

\begin{sidewaystable} 

\begin{longtable}{l|l|l|l}
Name & Ansatz $f(z)$ & Parameters of the basic equation& Comment\\
\hline\hline cosh I & $\frac{b}{1 + \lambda \cosh(\mu \: z)}$ &
$\alpha = \frac{\mu^2}{b^2} (1 - \lambda^2), \beta =
-\frac{\mu^2}{2
 b}, \gamma = \frac{\mu^2}{6}, \delta = \epsilon = 0$& special case of eq. (\ref{Out.f1})\\
\hline
sinh I & $\frac{b}{1 + \lambda \sinh(\mu \: z)}$ & $\alpha = \frac{\mu^2}{b^2} (1 + \lambda^2), \beta =
-\frac{\mu^2}{2
 b}, \gamma = \frac{\mu^2}{6}, \delta = \epsilon = 0$& special case of eq. (\ref{Out.f1})\\
\hline
cosh II & $\frac{b}{1 + \lambda \cosh^2(\mu \: z)}$ & $\alpha = \frac{4 \mu^2}{b^2} (1 + \lambda),
\beta = -\frac{\mu^2}{b} (2 + \lambda), \gamma = \frac{2 \mu^2}{3}, \delta = \epsilon = 0$
& special case of eq. (\ref{Out.f1})\\\hline
sinh II & $\frac{b}{1 + \lambda \sinh^2(\mu \: z)}$ & $\alpha = \frac{4 \mu^2}{b^2} (1 - \lambda),
\beta = -\frac{\mu^2}{b} (2 - \lambda), \gamma = \frac{2 \mu^2}{3}, \delta = \epsilon = 0$
& special case of eq. (\ref{Out.f1})\\\hline
cosh III & $\frac{b \cosh^2(\mu \: z)}{1 + \lambda \cosh^2(\mu \: z)}$
&$\alpha = \frac{4 \lambda^2 \mu^2 (1 + \lambda)}{b^2}, \beta = -\frac{\mu^2 (3 b \lambda^2 + 2 b \lambda)}{b^2}$
& special case of eq. (\ref{Out.f1})\\
$n = 1$ & & $\gamma = \frac{2 \mu^2 (3 b^2 \lambda + b^2)}{3 b^2}, \delta = -b \mu^2, \epsilon = 0$ & \\\hline
sinh III & $\frac{b \sinh^2(\mu \: z)}{1 + \lambda \sinh^2(\mu \: z)}$
&$\alpha = \frac{4 \lambda^2 \mu^2 (1 - \lambda)}{b^2}, \beta = -\frac{\mu^2 (-3 b \lambda^2 + 2 b \lambda)}{b^2}$
& special case of eq. (\ref{Out.f1})\\
$n = 1$ & & $\gamma = \frac{2 \mu^2 (-3 b^2 \lambda + b^2)}{3 b^2}, \delta = b \mu^2, \epsilon = 0$ & \\\hline
cosine & $\lambda \cos(\mu z)$ & $\alpha = \beta = \delta = 0, \gamma = -\frac{\mu^2}{6}, \epsilon = \lambda^2 \mu^2$
& special case of eq. (\ref{Out.f1})\\\hline
sine & $\lambda \sin(\mu z)$ & $\alpha = \beta = \delta = 0, \gamma = -\frac{\mu^2}{6}, \epsilon = \lambda^2 \mu^2$
& special case of eq. (\ref{Out.f1})\\\hline
tanh-sech & $\tanh(\mu z)$ & $\alpha = \mu^2, \beta = \delta = 0, \gamma = -\frac{\mu^2}{3}, \epsilon = \mu^2$
& special case of eq. (\ref{Out.f1})\\\hline
\caption{\label{TableA1} Relationship between the
approach outlined here and some ad-hoc methods given in \cite{Waz2006a}.}
\end{longtable}

\end{sidewaystable}

\section{Summary}\label{Con}

A method is proposed to obtain exact elliptic solutions to
NLWEEs. The method is a
generalization of some recently established ad-hoc methods
\cite{Waz2006a} (cf. tab. \ref{TableA1}) and, furthermore, includes conditions for periodic
and solitary solutions. These are given compactly in terms of the
invariants of the Weierstrass' elliptic function $\wp$. Periodic and solitary solutions
can be deduced systematically from the general solution (\ref{Out.f4}) and (\ref{BBM.f8a}), respectively. Thus, a
further promising ad-hoc method has been shown and, exemplary,
solutions to the BBM equation have been evaluated. A balancing principle has been evaluated that simplifies the application
of this method (cf. tab. \ref{Table1}). It seems that
this method is a useful and easily manageable tool with
interesting features (balancing principle, conditions for periodic and solitary wave
solutions, PDC) to find out exact solutions to nonlinear
differential equations, especially solitary ones. It may be advantageous that this quite general method
can lead to free parameters in the solution, as shown for the BBM equation.\\
It has recently been shown \cite{Schue2006} for the
Kadomtsev-Petviashvili equation and the nonlinear Schrödinger
equation that if $\triangle \neq 0$ and special conditions hold,
eq. (\ref{Out.f4}) can be rewritten in terms of Jacobian elliptic
functions and used as a start solution for a linear superposition
principle that enlarges the solution set of periodic solutions
\cite{38}. For the Novikov-Veselov equation it has been shown
previously \cite{Nic2006c}, that a 2-solitary wave solution can be
evaluated by linear superposition of two 1-solitary wave solutions
to the Korteweg-de Vries equation. These 1-solitary wave solutions
have been evaluated by the approach outlined above. Thus, this
approach can also be used as a basis to construct periodic
superposition solutions and multi-solitary wave solutions.


\ack

I would like to thank Prof. Wazwaz for sending me preprints of his interesting articles \cite{Waz2006a,Waz2006b}.
Furthermore,
I would like to thank Prof. H. W. Schürmann for helpful
discussions. The work was supported by the German Science
Foundation (DFG) (Graduate College 695 "Nonlinearities of optical
materials").

\end{document}